\documentclass[
preprint,
aps,prd,
nofootinbib,
superscriptaddress,
tightenlines,
amsmath,
amssymb
]{revtex4}
\usepackage{bm}
\usepackage{color,graphicx}
\usepackage{amsmath}
\usepackage{amssymb}

\begin{document}
\title{Easy as $\pi^o$: On the Interpretation of Recent Electroproduction Results} 
%\label{sec-5}

\author{Gary R.~Goldstein} 
\email{gary.goldstein@tufts.edu}
\affiliation{Department of Physics and Astronomy, Tufts University, Medford, MA 02155 USA.}
\author{J. Osvaldo Gonzalez Hernandez} 
\email{jog4m@virginia.edu}
\affiliation{Department of Physics, University of Virginia, Charlottesville, VA 22901, USA.}
\author{Simonetta Liuti} 
\email{sl4y@virginia.edu}
\affiliation{Department of Physics, University of Virginia, Charlottesville, VA 22901, USA.}

\begin{abstract}
Our original suggestion to investigate exclusive $\pi^o$ electroproduction  as a method
for extracting from data the tensor charge, transversity, and other quantities related to  chiral odd 
generalized parton distributions is further examined.   
We now explain the details of the process: {\it i)} the connection between the helicity description and the cartesian basis; {\it ii)} the dependence on the momentum transfer squared, $Q^2$,  and {\it iii)} the angular momentum, parity, and charge conjugation
constraints ($J^{PC}$ quantum numbers). 
\end{abstract}

\maketitle
\baselineskip 3.0ex

\section{Introduction}
The recent development of a theoretical framework for deeply virtual exclusive scattering processes using the concept of
Generalized Parton Distributions (GPDs) has opened vast opportunities for understanding and interpreting hadron structure {\it including spin}  within QCD. 
Among the many important consequences is the fact that differently from both inclusive and semi-inclusive processes, GPDs can in principle provide essential information for determining the missing component to the nucleon longitudinal spin sum rule, which is identified with orbital angular momentum. A complete description of nucleon structure requires, however,  also the transversity (chiral odd/quark helicity-flip) GPDs, $H_T(x, \xi, t)$, $E_T (x, \xi, t)$, $\widetilde{H}_T (x, \xi, t)$, and $\widetilde{E}_T (x, \xi, t)$ \cite{Diehl_01}. 
Just as in the case of the forward transversity structure function, $h_1$, the transversity GPDs are expected to be more elusive quantities, not easily determined experimentally. 
It was nevertheless suggested in Ref.\cite{AGL} that deeply virtual exclusive neutral pion electroproduction can provide a direct channel to measure chiral-odd GPDs so long as the helicity flip ($ \propto \gamma_5$) contribution to the quark-pion vertex is dominant. 
This idea was borne amidst contrast, based on the objection that the $\gamma_5$ coupling should be
subleading compared to the leading twist, chiral-even ($\propto \gamma_\mu\gamma_5$) contribution. It was subsequently, very recently, endorsed by QCD practitioners \cite{GolKro}.
Furthermore, experimental data from Jefferson Lab Hall B \cite{Kub} are now being interpreted in terms of chiral-odd GPDs. 
What caused this change and made the transversity-based interpretation be accepted? 

Within a QCD framework at leading order the scattering amplitude factors into a nucleon-parton correlator described by GPDs, and a hard scattering part which includes the pion Distribution Amplitude (DA). Keeping to a leading order description, the cross sections four-momentum squared dependences are $\sigma_L \approx {\cal O}(1/Q^6)$  and $\sigma_T \approx {\cal O}(1/Q^8)$, for the longitudinal and transverse virtual photon polarizations, respectively. It is therefore expected that in the deep inelastic region $\sigma_L$ will be dominant.
Experimental data from both Jefferson Lab \cite{HallA,Kub} and HERA \cite{H1ZEUS} on meson electroproduction show exceedingly larger than expected  transverse components of the cross section. Clearly, an explanation lies beyond the reach of the leading order collinear factorization that was first put forth, and more interesting dynamics might be involved. 
In Ref.\cite{AGL} it was noted that sufficiently large transverse cross sections can be produced in $\pi^o$ electroproduction  provided the chiral odd coupling is  adopted. 
%[What happens to factorization. What happens to $\rho$ production.] 
Since all available data are in the few GeV kinematical region it should not be surprising to find such higher twist terms to be present.      
What makes it more difficult to accept is that the longitudinal components of the cross section should simultaneously be suppressed.  
Also, given the trend of the HERA data, why is this canonical higher twist effect seemingly not disappearing at even larger $Q^2$ values? In order to answer this question, in this paper we carefully address, one by one, the issues of the construction of helicity amplitudes and their connection with the cartesian basis (Section \ref{sec2}), the $Q^2$-dependence (Section \ref{sec3}), the role of $t$-channel spin, parity, charge conjugation and their connection to the GPDs crossing symmetry properties (Section \ref{sec4}), finally giving our conclusions in Section \ref{sec5}. 

%%%%
%%%
%%% FIGURE 1
%%%
\begin{figure}
\includegraphics[width=9.cm]{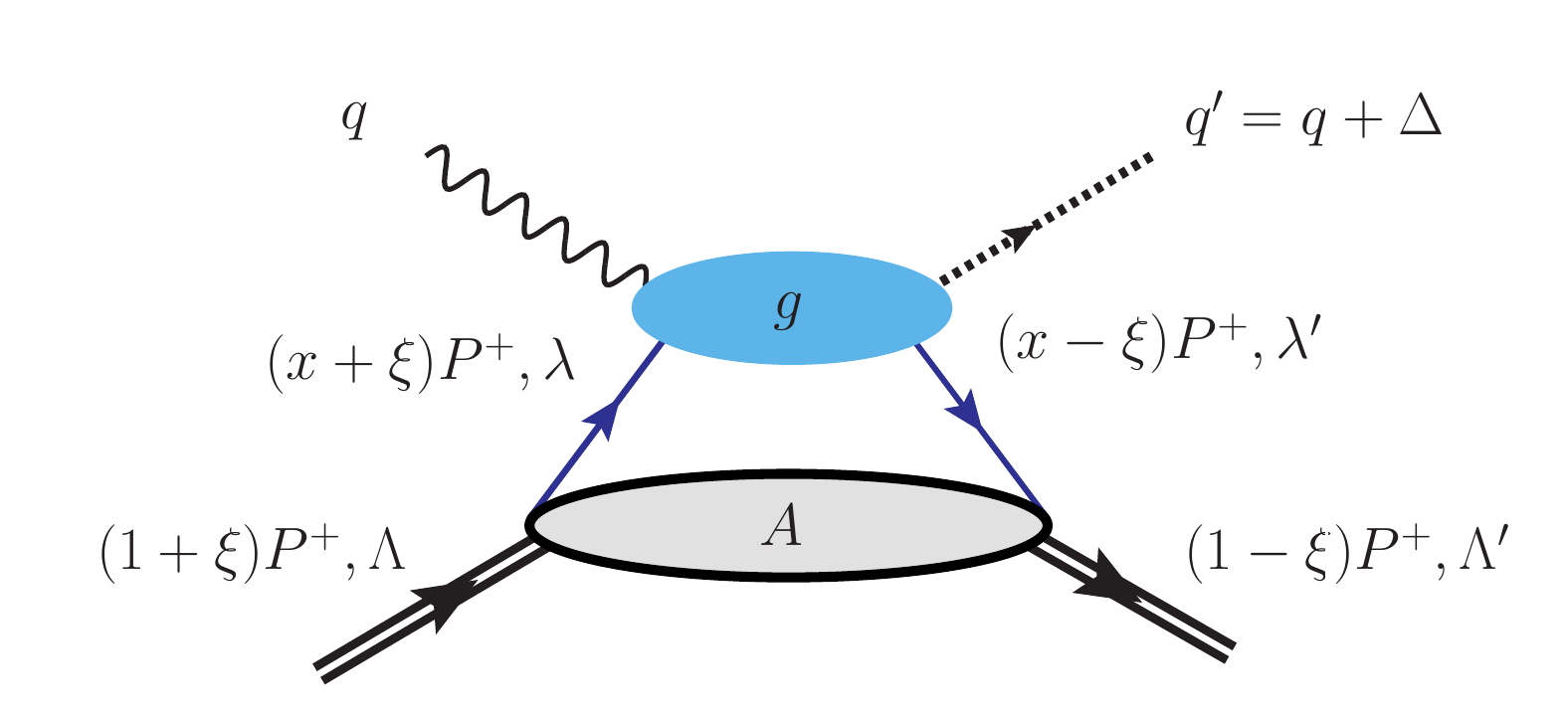}
\includegraphics[width=7.cm]{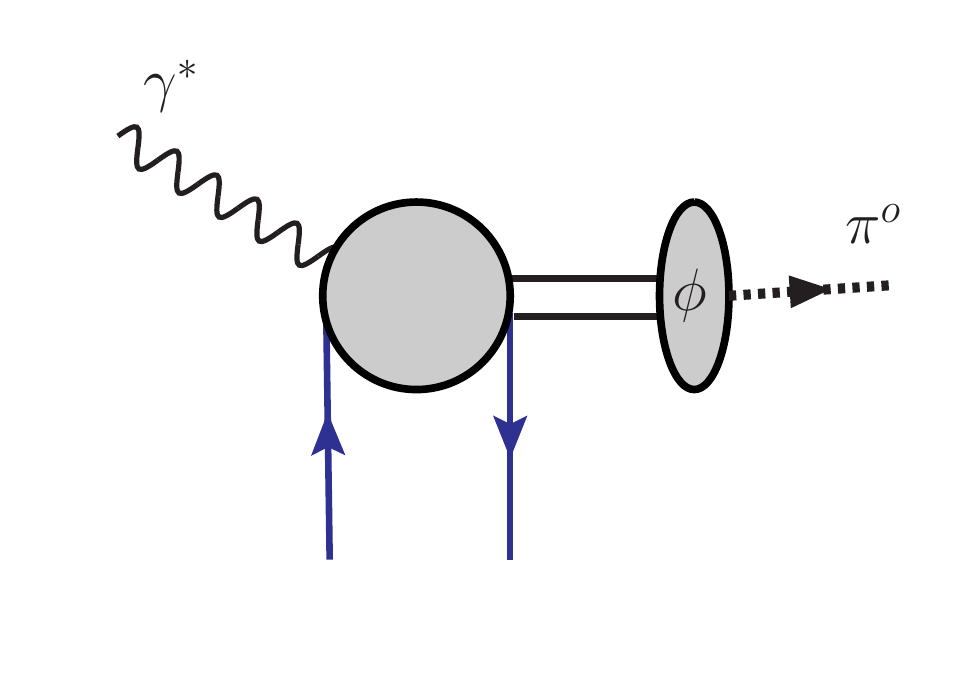}
\caption{Left: Leading order amplitude for DVMP, $\gamma^* + P \rightarrow M +P^\prime$. Notice that differently from Refs. \cite{GGL_even,GGL_odd}, we adopt the symmetric scheme of kinematics where $\overline{P}=(P+P')/2$ (see Ref.\cite{BelRad} for a review). 
Right: Hard scattering contribution to the DVMP process, where $\phi$ is the outgoing meson distribution amplitude. 
%Crossed diagrams are not shown in the figure.
}
\label{fig1}
\end{figure}

%%%%%%%%%%%%%%%%%%%%%%%%%%%%%%%%%%%%%%%%%%%%%%%%%%%%%%%%%%%%%%%%%%%%%%%%%%%%%%%%%%%%%
%%%%%%%%%%%%%%%%%%%%%%%%%%%%%%%%%%%%%%%%%%%%%%%%%%%%%%%%%%%%%%%%%%%%%%%%%%%%%%%%%%%%%
%%%%%%%%%%%%%%%%%%%%%%%%%%%%%%%%%%%%%%%%%%%%%%%%%%%%%%%%%%%%%%%%%%%%%%%%%%%%%%%%%%%%%
%%%%%%
%%%%%% SECTION 2
\section{Cross Section in terms of GPDs}
\label{sec2}
%Normalization: how we fix it
%Composition of cross section in terms of GPDs
The amplitude for Deeply Virtual Meson Production (DVMP) off a proton target is written  in a QCD factorized picture, using the helicity formalism, as the following convolution over the quark momentum components \cite{AGL,GGL_even}, 
\begin{eqnarray}
f_{\Lambda_\gamma,\Lambda;0,\Lambda^\prime}(\xi,t) & = & \sum_{\lambda,\lambda^\prime} 
\int dx d^2k_\perp \; g_{\Lambda_\gamma,\lambda;0,\lambda^\prime} (x, k_\perp, \xi, t)  \,
A_{\Lambda^\prime,\lambda^\prime;\Lambda,\lambda}(x, k_\perp, \xi,t)
\label{facto}
\end{eqnarray}
where the variables $x, \xi, t$ are $x=(k^+ + k^{\prime \, +})/(P^+ + P^{\prime \, +})$, $\xi=\Delta^+/(P^+ + P^{\prime \, +})$, $t=\Delta^2$; the particles momenta and helicities,  along with the hard scattering amplitude, $g_{\Lambda_\gamma,\lambda;0,\lambda^\prime}$, and the quark-proton scattering amplitude, $A_{\Lambda^\prime,\lambda^\prime;\Lambda,\lambda}$, are displayed in Fig.\ref{fig1}. The $Q^2$ dependence is omitted for ease of presentation.
% are the matrix elements for the helicity dependent quark operators and proton states.  
%Eq.(\ref{facto}) describes both the chiral even $(\lambda = \lambda^\prime)$ and chiral odd $(\lambda \neq \lambda^\prime) $ contributions. 

In our approach the chiral even contribution is sub-leading (see Section 3).
As explained in detail in Refs.\cite{AGL,GGL_odd}, for the chiral odd contribution to $\pi^o$ electroproduction only $g_{1+,0-} \equiv g_T$, and $g_{0+,0-} \equiv g_L$, are different from zero. Furthermore $g_L$ is suppressed at ${\cal O}(k'_\perp/Q)$. Their expressions are
%\footnote{These expressions differ from the ones in Ref.\cite{GGL_odd} because the partons' spinor normalizations.}  
\begin{eqnarray}
\label{gT_odd}
g_T^{A,V} & = &  g_\pi^{odd}(Q) \left[\frac{1}{x-\xi + i \epsilon} -  \frac{1}{x + \xi - i \epsilon}  \right] = g_{\pi \, odd}^{A,V}(Q) \; C^- \\
\label{gL_odd}
g_L^{A,V} & = &   g_\pi^{odd}(Q)  \sqrt{\frac{t_o-t}{Q^2}} \left[  \frac{1}{x-\xi + i \epsilon} -  \frac{1}{x +\xi - i \epsilon} \right] =  
%\frac{ g_\pi^{odd}(Q) \zeta k_\perp^\prime}{\sqrt{Q^2}(X-\zeta)} \; C^+,
g_{\pi \, odd}^{A,V}(Q)  \sqrt{\frac{t_o-t}{Q^2}}   \; C^-,
\end{eqnarray}
where $g_{\pi \, odd}^{A,V}(Q)$ describe the pion vertex. The labels $A,V$ describe axial-vector and vector exchanges as we will explain in Section \ref{sec3}. 
Notice that the coefficients $C^-$ from the quark propagator is even under crossing.
Using the allowed hard scattering amplitudes we obtain the following six independent amplitudes, of which four are for the transverse photon,
\begin{subequations}
\label{f_T}
\begin{eqnarray}
\label{f1}
f_1 = f_{1+,0+} &  = & g_{1+,0-} \otimes A_{+-,++}  \\
\label{f2}
f_2 = f_{1+,0-} &  = &  g_{1+,0-}  \otimes  A_{--,++}     \\
\label{f3}
f_3  = f_{1-,0+} & = &  g_{1+,0-}  \otimes  A_{+-,-+}    \\  
\label{f4}
f_4  = f_{1-,0-} & = &   g_{1+,0-}   \otimes  A_{--,-+},  
\end{eqnarray}
\label{chiral_odd_T}
\end{subequations}
and two for the longitudinal photon,
\begin{subequations}
\label{f_L}
\begin{eqnarray}
\label{f5}
f_5 = f_{0+,0-} &  = & g_{0+,0-}  \otimes  A_{--,++}    \\
\label{f6}
f_6 = f_{0+,0+} &  = & g_{0+,0-}  \otimes  A_{+-,++}, 
\label{chiral_odd_L}
\end{eqnarray}
\end{subequations}
The $\pi^o$ electroproduction cross section is given by,
\begin{eqnarray} 
\label{xs}
\frac{d^4\sigma}{d\Omega d\epsilon_2 d\phi dt} & = &\Gamma \left\{ \frac{d\sigma_T}{dt} + \epsilon_L \frac{d\sigma_L}{dt} + \epsilon \cos 2\phi \frac{d\sigma_{TT}}{dt} 
+ \sqrt{2\epsilon_L(1+\epsilon)} \cos \phi \frac{d\sigma_{LT}}{dt}  \right. \nonumber \\
& + & \left.
h  \, \sqrt{2\epsilon_L(1-\epsilon)} \, \frac{d\sigma_{L^\prime T}}{dt} \sin \phi \right\}, 
\end{eqnarray}
where, $\Gamma=(\alpha/2\pi^2) (k_e'/k_e)(k_\gamma/Q^2) /(1-\epsilon)$, with $k_e, k_e'$ being the initial and final electron energies, $k_\gamma$ the real photon equivalent energy in the lab frame, $h=\pm1$ for the electron beam polarization, $\epsilon$ and $\epsilon_L=Q^2/\nu^2 \epsilon$, are the transverse and longitudinal polarization fractions, respectively \cite{DreTia,DonRas}, and the cross sections terms read,
\begin{subequations}
\label{xsecs}
\begin{eqnarray}
%\label{dsigT}
\frac{d\sigma_T}{dt} & = & \mathcal{N}  \, \left( \mid f_1 \mid^2 + \mid f_2 \mid^2 + \mid f_3 \mid^2 +
\mid f_4 \mid^2 \right)  
\\
%\label{dsigL}
\frac{d\sigma_L}{dt} & = & \mathcal{N} \,  \left( \mid f_5 \mid^2 + \mid f_6 \mid^2 \right),  \\
\label{dsigTT}
\frac{d\sigma_{TT}}{dt} & = & 2 \, \mathcal{N} \,    \Re e \left( f_1^*f_4 - f_2^* f_3 \right). \\
%%%%%
%\label{dsigLT}
\frac{d\sigma_{LT}}{dt} & = & 2 \, \mathcal{N} \, 
\Re e \left[ f_5^* (f_2 + f_3) + f_6^* (f_1 - f_4) \right]. \\
%%%%
%\label{dsigLTp}
\frac{d\sigma_{LT^\prime}}{dt} & = & 2 \, \mathcal{N}  \, 
\Im m  \left[ f_5^* (f_2 + f_3)  + f_6^* (f_1 - f_4) \right]
\end{eqnarray}
\end{subequations}
with $\mathcal{N}=(\pi/2)/[s (s-M^2)]$.  
In order to understand how the chiral odd GPDs enter the helicity structure we need to make a connection between the helicity amplitudes in Eqs.(\ref{f_T},\ref{f_L}), and the four Lorentz structures for the hadronic tensor first introduced in Ref.\cite{Diehl_01} (cartesian basis),
\begin{eqnarray}
\label{CFF_odd}
&& \epsilon^\mu_T T_\mu^{\Lambda \Lambda^\prime}  =    e_q \int_{-1}^1  dx \:   
\frac{g_{T}}{2 \overline{P}^+} \; {\overline{U}(P',\Lambda')}\left[ i \sigma^{+i} H_T^q(x,\xi,t) +   
\frac{\gamma^+ \Delta^i - \Delta^+ \gamma^i}{2M} E_T^q(x,\xi,t) \right.  \nonumber \\
&& \left. \frac{\overline{P}^+ \Delta^i - \Delta^+ \overline{P}^i}{M^2}  \widetilde{H}_T^q(x,\xi,t) +
\frac{\gamma^+ \overline{P}^i - \overline{P}^+ \gamma^i}{2M} \widetilde{E}_T^q(x,\xi,t) \right] U(P,\Lambda),
\end{eqnarray}
where $i=1,2$, and $\epsilon^\mu_T$ is the transverse photon polarization vector. An explicit calculation of each factor multiplying the GPDs in Eq.(\ref{CFF_odd}) yields,
\begin{subequations}
\begin{eqnarray}
h_T^{i, \, \Lambda \Lambda^\prime} & = & \overline{U}(P^\prime,\Lambda^\prime) i \sigma^{+i}  U(P,\Lambda)  =   f  \;   \frac{1-\xi}{1+\xi} (\Lambda  \delta_{i1} + i \delta_{i2} ) \delta_{\Lambda,-\Lambda'}\\ 
%%%% H_T tilde
\tilde{h}_T^{i, \, \Lambda \Lambda^\prime} &=& \overline{U}(P^\prime,\Lambda^\prime) \frac{\overline{P}^+ \Delta^i - \Delta^+ \overline{P}^i}{M^2}   U(P,\Lambda)  = f 
\left[  \frac{1}{1+\xi} \frac{\Delta_i}{M} \delta_{\Lambda, \Lambda'}  + (\Lambda \Delta_1+ i  \Delta_2) \frac{\Delta_i}{2M^2} \delta_{\Lambda,- \Lambda'}  \right] \nonumber \\
&& \\
%%%% E_T
e_T^{i, \, \Lambda \Lambda^\prime} &=&   \overline{U}(P^\prime,\Lambda^\prime)   \frac{\gamma^+ \Delta^i - \Delta^+ \gamma^i}{2M}  U(P,\Lambda)  = \nonumber \\
&& f \left[ \frac{1}{1+\xi} \left( \frac{\Delta_i}{2M} + i \Lambda \, \xi \, \epsilon^{03ji} \frac{\Delta_j}{2M} \right)  \delta_{\Lambda \Lambda'} +  
\frac{\xi^2}{(1+\xi)^2} (\Lambda \delta_{i1} + i \delta_{i2}) \delta_{\Lambda,- \Lambda'}  \right ] 
 \\
%%%% E tilde
\tilde{e}_T^{\, i, \, \Lambda \Lambda^\prime} &=& \overline{U}(P^\prime,\Lambda^\prime)   \frac{\gamma^+ \overline{P}^i - \overline{P}^+ \gamma^i}{M}  U(P,\Lambda)   =  \nonumber \\
&&  f \left[  \frac{1}{1+\xi} \left( \xi \frac{\Delta_i}{2M} + i \, \Lambda \, \epsilon^{03ji} \frac{\Delta_j}{2M} \right) \delta_{\Lambda \Lambda'}   +
\frac{\xi}{(1+\xi)^2} ( \Lambda \delta_{i1} + i \delta_{i2} ) \delta_{\Lambda,- \Lambda'}  \right ]
\end{eqnarray}
\end{subequations}
with,
$f = 1/[2\sqrt{1-\xi^2} ]$.
We now  take in the expressions above the combinations,
\begin{subequations}
\begin{eqnarray}
 f_T^{\Lambda \Lambda'} & = & f_T^{1, \Lambda \Lambda'} + i f_T^{2, \Lambda \Lambda'}   \; \; \; \Lambda=\Lambda^\prime \\
 f_T^{\Lambda \Lambda'} & = & f_T^{1, \Lambda \Lambda'} - i f_T^{2, \Lambda \Lambda'}   \; \; \; \Lambda \neq \Lambda^\prime,
 \end{eqnarray}
\end{subequations}
where  $f_T=h_T,e_T,\tilde{h}_T,\tilde{e}_T$, and without loss of generality consider all vectors lying in the $x$-$z$ plane with \[\mid \Delta_\perp \mid \equiv \Delta_1 = \sqrt{t_0-t}\sqrt{(1-\xi)/(1+\xi)},\]

\vspace{0.3cm} \noindent
$(\Lambda,\Lambda^\prime) = ++, --$
\begin{subequations}
\label{coeff++}
\begin{eqnarray}
%%%% H_T
h_T^{1, \, \Lambda \Lambda^\prime} +  i h_T^{2, \Lambda \Lambda'} & = & 0  \\ 
%%%% H_T tilde
\tilde{h}_T^{1, \, \Lambda \Lambda^\prime} + i  \tilde{h}_T^{2, \, \Lambda \Lambda^\prime}  &=&  f 
\left[  \frac{1}{1+\xi} \frac{\Delta_1 + i \Delta_2}{M}     \right]  
 =  \frac{\sqrt{t_0-t}}{M(1+\xi)^2}     \\
%%%% E_T
e_T^{1, \, \Lambda \Lambda^\prime} + i e_T^{2, \, \Lambda \Lambda^\prime} &=&   
 f \left[ \frac{1}{1+\xi} \left( \frac{\Delta_1 + i  \Delta_2}{2M}  + i \Lambda \, \xi \, \frac{\Delta_2 + i \Delta_1}{2M} \right)   
  \right ]  
 =     \frac{\sqrt{t_0-t}}{2M(1+\xi)^2} (1 \mp \xi) \nonumber \\
 && \\
%%%% E tilde
\tilde{e}_T^{\, 1, \, \Lambda \Lambda^\prime} + i \tilde{e}_T^{\, 2, \, \Lambda \Lambda^\prime} &=&  f \left[  \frac{1}{1+\xi} \left( \xi \frac{\Delta_1 + i \Delta_2 }{2M} + i \, \Lambda \frac{\Delta_2 + i \Delta_1}{2M} \right)     \right ] 
 =  \frac{\sqrt{t_0-t}}{2M(1+\xi)^2} (\xi \mp 1) \nonumber \\
\end{eqnarray}
\end{subequations}
%%%%%%
%%%%%%

\noindent \vspace{0.3cm}
$(\Lambda,\Lambda^\prime) = -+$
\begin{subequations}
\label{coeff+-}
\begin{eqnarray}
%%%% H_T
h_T^{1, \, \Lambda \Lambda^\prime} -  i h_T^{2, \Lambda \Lambda'} & = &   0  \\
%%%% H_T tilde
\tilde{h}_T^{1, \, \Lambda \Lambda^\prime} - i  \tilde{h}_T^{2, \, \Lambda \Lambda^\prime}  &=&  2 f 
   \frac{\Delta_i}{2M^2}  
    = -  \frac{{t_0-t}}{M^2} \frac{\sqrt{1-\xi^2}}{(1+\xi)^2} \\
%%%% E_T
e_T^{1, \, \Lambda \Lambda^\prime} - i e_T^{2, \, \Lambda \Lambda^\prime} &=& 0 \\
%%%% E tilde
 \tilde{e}_T^{\, 1, \, \Lambda \Lambda^\prime} - i \tilde{e}_T^{\, 2, \, \Lambda \Lambda^\prime} &=&  0 
\end{eqnarray}
\end{subequations}

\noindent \vspace{0.3cm}
$(\Lambda,\Lambda^\prime) = +-$
\begin{subequations}
\label{coeff-+}
\begin{eqnarray}
%%%% H_T
h_T^{1, \, \Lambda \Lambda^\prime} -  i h_T^{2, \Lambda \Lambda'} & = &   2 \, f  \;   \frac{1-\xi}{1+\xi}  
 =   2 \frac{\sqrt{1-\xi^2}}{(1+\xi)^2} \\
%%%% H_T tilde
\tilde{h}_T^{1, \, \Lambda \Lambda^\prime} - i  \tilde{h}_T^{2, \, \Lambda \Lambda^\prime}  &=&  2 f 
   \frac{\Delta_i}{2M^2} 
    =   \frac{{t_0-t}}{M^2} \frac{\sqrt{1-\xi^2}}{(1+\xi)^2} \\
%%%% E_T
e_T^{1, \, \Lambda \Lambda^\prime} - i e_T^{2, \, \Lambda \Lambda^\prime} &=& 2   f  \frac{\xi^2}{(1+\xi)^2}  
 =   \frac{\sqrt{1-\xi^2}}{1+\xi^2}  \,  \frac{\xi^2}{(1-\xi)^2}  \\
%%%% E tilde
 \tilde{e}_T^{\, 1, \, \Lambda \Lambda^\prime} - i \tilde{e}_T^{\, 2, \, \Lambda \Lambda^\prime} &=&  2 f  \frac{\xi}{(1+\xi)^2} 
= \frac{\sqrt{1-\xi^2}}{1+\xi^2}  \, \,  \frac{\xi}{(1-\xi)^2} .
\end{eqnarray}
\end{subequations}
Eqs.(\ref{coeff++},\ref{coeff+-},\ref{coeff-+}) provide the connection between the cartesian and helicity bases: they give the coefficients multiplying the GPDs which enter each one of the helicity amplitudes, $f_1 (\Lambda,\Lambda^\prime=++), f_2 (\Lambda,\Lambda^\prime=+-), f_3 (\Lambda,\Lambda^\prime=-+), f_4 (\Lambda,\Lambda^\prime=--), f_5 (\Lambda,\Lambda^\prime=++), f_6 (\Lambda,\Lambda^\prime=+-)$.   

The GPD content of each amplitude therefore is,
%We can now write the helicity amplitudes in terms of GPDs,
\begin{subequations}
\label{f_amps_0}
\begin{eqnarray}
f_1 & \rightarrow &  \frac{\sqrt{t_0-t}}{2M(1+\xi)^2} \left[ 2\widetilde{\cal H}_T  + (1-\xi)  \left({\cal E}_T - \widetilde{\cal E}_T \right) \right] \\
%%%
f_2 & \rightarrow &     \frac{\sqrt{1-\xi^2}}{(1+\xi)^2}  \left[ {\cal H}_ T + \frac{t_0-t}{4M^2} \widetilde{\cal H}_T      
+ \frac{\xi^2}{1-\xi^2}  {\cal E}_T  + \frac{\xi}{1-\xi^2} \widetilde{\cal E}_T \right]  \\
%%%
\nonumber \\ 
f_3  & \rightarrow &     \, -  \frac{\sqrt{1-\xi^2}}{(1+\xi)^2}  \,  \frac{t_0-t}{4M^2} \, \widetilde{\cal H}_T  
 \\
%%%
f_4 & \rightarrow &    \frac{\sqrt{t_0-t}}{2M(1+\xi)^2}  \,  \left[  2\widetilde{\cal H}_ T + (1+\xi) \left( {\cal E}_T +
 \widetilde{\cal E}_T \right) \right], 
\end{eqnarray} 
\end{subequations}
%$\xi=x_{Bj}/(2-x_{Bj})$, 
where we write,
\[ {\cal F}_T(\xi,t,Q^2) = \int_{-1}^1 dx \; C^- \, F_T(x,\xi,t,Q^2) \; \; \; \; \;  F_T \equiv {\cal H}_T, {\cal E}_T, \widetilde{\cal H}_ T, \widetilde{\cal E}_ T. \]  
The quark flavor content  of the quark-proton helicity amplitudes for $\pi^o$ electroproduction is, 
\[A_{\Lambda^\prime,\lambda^\prime;\Lambda,\lambda} = e_uA^u_{\Lambda^\prime,\lambda^\prime;\Lambda,\lambda}-e_dA^d_{\Lambda^\prime,\lambda^\prime;\Lambda,\lambda}. \]
The amplitudes for longitudinal photon polarization, $f_5$ and $f_6$ are obtained similarly, by working out  Eqs.(\ref{f5},\ref{f6}) (details are given in Ref.\cite{GGL_odd}).

We now proceed to discuss the $Q^2$ dependence of the process. Our description of the hard scattering part, Fig.\ref{fig1}, yields $Q^2$ dependent coefficients appearing in the 
various helicity amplitudes, 
\begin{subequations}
\label{f_amps_1}
\begin{eqnarray}
f_1 &= &  g_{\pi \, odd}^V(Q) \frac{\sqrt{t_0-t}}{2M(1+\xi)^2} \left[ 2\widetilde{\cal H}_T  + (1-\xi)  \left({\cal E}_T - \widetilde{\cal E}_T \right) \right] \\
%%%
f_2 & = &    (g_{\pi, \, odd}^V(Q) + g_{\pi, \, odd}^A(Q))  \,  \frac{\sqrt{1-\xi^2}}{(1+\xi)^2}  \left[ {\cal H}_ T + \frac{t_0-t}{4M^2} \widetilde{\cal H}_T      
+ \frac{\xi^2}{1-\xi^2}  {\cal E}_T  + \frac{\xi}{1-\xi^2} \widetilde{\cal E}_T \right]  \\
%%%
\nonumber \\ 
f_3  & = &   (g_{\pi, \, odd}^V(Q) - g_{\pi, \, odd}^A(Q))  \,   \frac{\sqrt{1-\xi^2}}{(1+\xi)^2}  \,  \frac{t_0-t}{4M^2} \, \widetilde{\cal H}_T  
 \\
%%%
f_4 & = &  g_{\pi \, odd}^V(Q) \,  \frac{\sqrt{t_0-t}}{2M(1+\xi)^2}  \,  \left[  2\widetilde{\cal H}_ T + (1+\xi) \left( {\cal E}_T +
 \widetilde{\cal E}_T \right) \right]. 
\end{eqnarray} 
\end{subequations}
These coefficients are derived accordingly to the values of the $-t$-channel $J^{PC}$  quantum numbers for the process. They describe an angular momentum dependent formula that we discuss in detail in the following Section.

From our discussion so far, it clearly appears how the interpretation of exclusive meson electroproduction, even assuming the validity of factorization, depends on several components which make the sensitivity of the process to transversity and related quantities hard to disentangle.  
For a more streamlined physical interpretation of the cross section terms, one can define leading order expressions, {\it i.e.}  take the dominant terms in $Q^2$ in Eqs.(\ref{f_amps_1}), and small $\xi$ and $t$ in Eqs.(\ref{f_amps_0},\ref{f_amps_1}). Then the dominant GPDs contributions to Eqs.(\ref{xsecs}) can be singled out as, 
\begin{eqnarray}
\frac{ d \sigma_T }{dt } & \approx & {\cal N}  \;  \left[
\mid {\cal H}_T \mid^2 + \,  \tau \left( \mid  \overline{\cal E}_T  \mid^2 + \mid  \widetilde{\cal E}_T  \mid^2 \right)  \right]   \\
\frac{ d \sigma_L }{dt } & \approx & {\cal N}  \; \frac{2M^2  \tau}{Q^2}  \mid {\cal H}_T \mid^2 \\
\frac{ d \sigma_{TT} }{dt } & \approx   &  {\cal N}     \;  \tau
\left[ \mid \overline{\cal E}_T \mid^2 -\mid  \widetilde{\cal E}_T \mid^2   + \;
\Re e  {\cal H}_T  \,  \frac{\Re e (\overline{\cal E}_T   - {\cal E}_T)}{2}  + \Im m  {\cal H}_T  \,  \frac{\Im m  (\overline{\cal E}_T   -  {\cal E}_T)}{2}     \right]  \nonumber \\
&& \\
\frac{ d \sigma_{LT} }{dt } & \approx &  {\cal N}    \; 2 \, \sqrt{ \frac{2M^2  \, \tau}{Q^2}}  \mid {\cal H}_T \mid^2 \\
\frac{ d \sigma_{L^\prime T} }{dt } & \approx & {\cal N} \; \tau \, \sqrt{\frac{2 M^2 \, \tau}{Q^2} } \;  \left[  \Re e  {\cal H}_T  \,   \frac{\Im m (\overline{\cal E}_T   -  {\cal E}_T)}{2}   -   \Im m  {\cal H}_T  \,   \frac{\Re e   (\overline{\cal E}_T   -  {\cal E}_T)}{2}  \right]
\end{eqnarray}
where $\tau= (t_o-t)/2M^2$, and we have redefined $\overline{\cal E}_T = 2 {\cal H}_T + {\cal E}_T$ according to \cite{Bur2}. 

Very little is known on the size and overall behavior of the chiral odd GPDs, besides that $H_T$ becomes the transversity structure function, $h_1$, in the forward limit,  $\overline{E}_T$'s first moment is  the proton's transverse anomalous magnetic moment \cite{Bur2},  and $\widetilde{E}_T$'s first moment is null \cite{Diehl_01}. 
To evaluate the chiral odd GPDs in Ref. \cite{GGL_odd} we propose a method that, by using Parity transformations in a spectator picture, allows us to write them as linear combinations of the better determined chiral even GPDs \cite{GGL_even}. 
%Here we present results using such a procedure, while we postpone its detailed description to Ref.\cite{GGL_odd}.  

Based on our analysis we expect the following behaviors to approximately appear in the data:

\noindent {\it i)} The order of magnitude of the various terms approximately follows a sequence determined by the inverse powers of $Q$ and the powers of $\sqrt{t_o-t}$: $d \sigma_T/dt \geq d \sigma_{TT}/dt \geq d \sigma_{LT/L'T}/dt \geq d \sigma_L/dt; $

\noindent {\it ii)} $d \sigma_T/dt$ is dominated by ${\cal H}_T$ at small $t$, and governed by the interplay of  ${\cal H}_T$ and $\overline{\cal E}_T$ at larger $t$; 

\noindent {\it iii)} $d \sigma_L/dt$ and $d \sigma_{LT}/dt$ are directly sensitive to ${\cal H_T}$; 

\noindent {\it iv)} $d \sigma_{TT}/dt$ and $d \sigma_{L'T}/dt$ contain a mixture of GPDs. They will play an important role in 
singling out the less known terms, $\overline{E}_T$, $E_T$, and $\widetilde{E}_T$. 

\begin{figure}
\includegraphics[width=7.5cm]{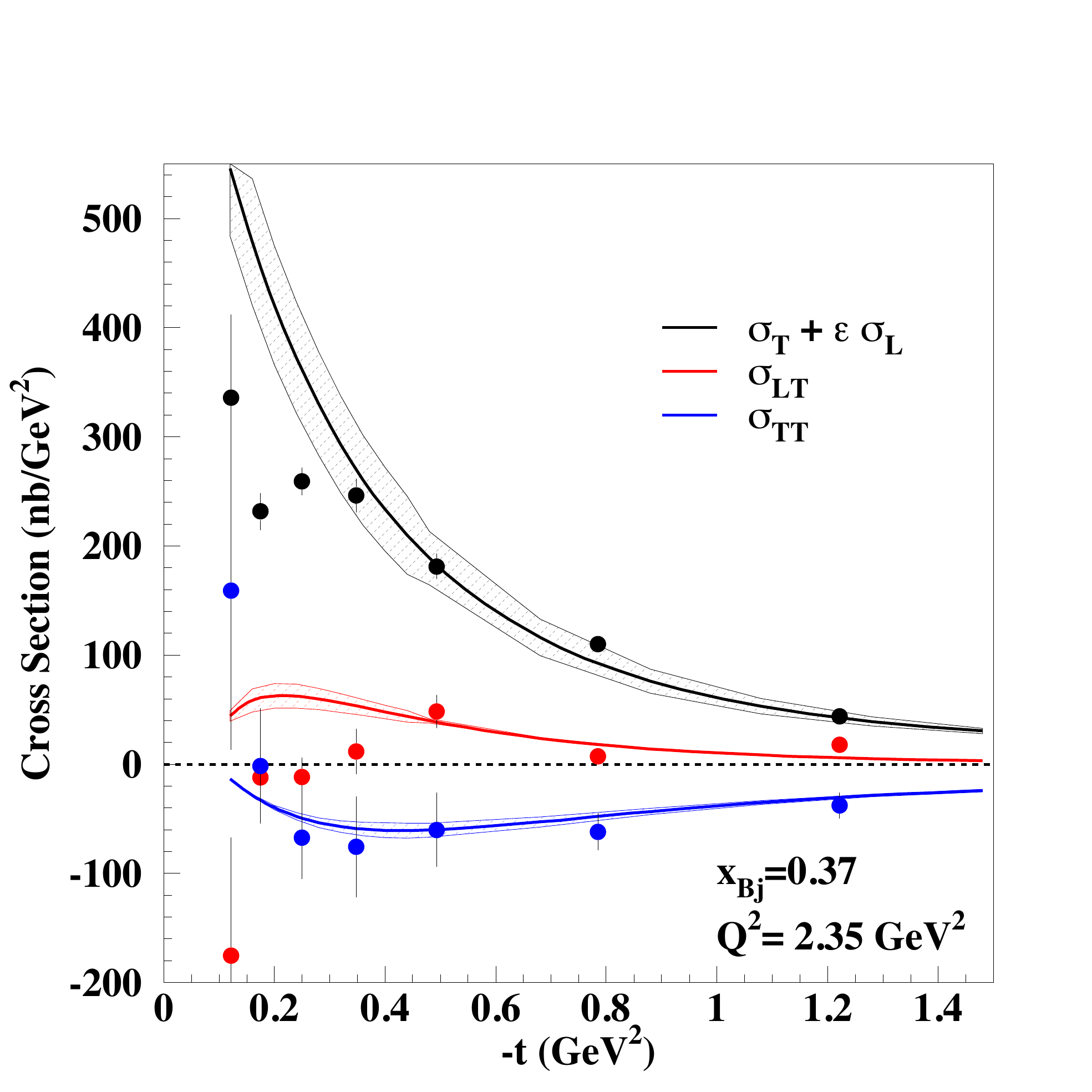}
\caption{(color online) $d \sigma_T/dt + \epsilon_L d \sigma_L/dt$, $d \sigma_{TT}/dt$, and $d \sigma_{LT}$ calculated using the physically motivated parametrization of Ref.\cite{GGL_odd}, plotted vs. $-t$ for $x_{Bj}=0.37$, and $Q^2=2.35$ GeV$^2$, along with data from Hall B \cite{Kub}. The hatched areas represent the theoretical errors for the parametrization, which are mostly originating from the fit to the nucleon form factors \cite{GGL_even,GGL_odd}.}
\label{fig2}
\end{figure}

%%%%%%%
\vspace{0.3cm}
\noindent The interplay of the various GPDs can already be seen by comparing to the Hall B data \cite{Kub} shown in  in Fig.\ref{fig2}. 
%%%%
%%%
%%% FIGURE 2
%%%
%%%%%%
One can see, for instance, that the ordering predicted in {\it i)} is followed, and that  $d \sigma_T/dt$ exhibits a form factor-like fall  off of ${\cal H}_T$ with $-t$. 

\section{Pseudo-scalar meson coupling}
\label{sec3}
%%%%% Two different couplings
% The two different couplings for $\pi^0$ electroproduction at high energy and the $Q^2$ dependence}
%
The hard part of $\gamma^* p \rightarrow \pi^o p^\prime$ involves the $\gamma^* + u(d) \rightarrow \pi^o + u(d)$ amplitudes (Fig.\ref{fig1}). 
%%%% Short Intro to the topic
In the longitudinal $\gamma^*$ case the near collinear limit ($-t<<Q^2$) admits at leading order, amplitudes in which the quark does not flip helicity. 
The $\pi^0$'s non-flip quark vertex is via a $\gamma^\mu \gamma^5$ coupling, which corresponds to a twist-2 contribution. 
The non-flip transverse contribution is suppressed -- twist-4. 
%The axial vector must contract with the $\pi^0$ momentum. 
For transverse $\gamma^*$ the quark can also flip helicity in the near collinear limit. 
This is accomplished through a vertex with $\gamma^5$ coupling giving the same $Q^2$ dependence as in the non-flip case. 
%By itself such a coupling is twist-3. 
%However, because of the larger -- twist-4 -- suppression in the non-flip case, for transverse $\gamma^*$ 
However, based on the role of $J^{PC}$ quantum numbers (Section 3), we argue that these transverse amplitudes will be 
dominating DVMP cross sections in the multi-GeV region.
%%%% End short Intro

The $\pi^o$ vertex  is described in terms of Distribution Amplitudes (DAs) as follows \cite{Huang,BenFel}, 
\begin{eqnarray}
{\cal P} & = & K f_\pi \left\{ \gamma_5  \not\!{q}^\prime \phi_\pi(\tau) + \gamma_5 \mu_\pi  \phi_\pi^{(3)} (\tau) \right\}
\label{pi_coupling}
\end{eqnarray}
where $f_\pi$ is the pion coupling,  $\mu_\pi$ is a mass term that can {\it e.g.} be estimated from the gluon condensate, $\phi_\pi(\tau)$ and $\phi_\pi^{(3)} (\tau)$, $\tau$ being the longitudinal momentum fraction, are the twist-2 and twist-3 pion DAs, respectively describing the chiral even and chiral odd processes.
We now evaluate the $Q^2$ dependence of the process. 
In Ref.\cite{GuichonVdh}, by assuming a simple collinear one gluon exchange mechanism it was shown that for the chiral even case,
\begin{eqnarray}
\label{gL_coll}
g_{0,+; 0,+} &  \approx  & \frac{1}{Q} \int d \tau \frac{\phi_\pi(\tau)}{\tau}  C^- 
\Rightarrow \frac{d\sigma_L^{even}}{dt} \propto \frac{1}{Q^6} \\
\label{gT_coll}
g_{1,+; 0,+} &   \approx  & \frac{1}{Q^2} \int d \tau \frac{\phi_\pi(\tau)}{\tau}  C^- 
\Rightarrow \frac{d\sigma_T^{even}}{dt} \propto \frac{1}{Q^8}. 
\end{eqnarray}
%where to evaluate the cross section dependence one uses Eq.(\ref{xsecs}), with ${\cal N} \propto 1/Q^4$. 
%and we disregarded scaling violating type contributions. 
By inserting Eq.(\ref{pi_coupling}) in Eqs.(\ref{gT_odd},\ref{gL_odd}), based on the same mechanism, in the chiral-odd case one has  
\[  g_{0+,0-} \approx \frac{d\sigma_L^{odd}}{dt} \propto \frac{1}{Q^{10}},  \; \; \;  g_{1+,0-} \approx \frac{d\sigma_T^{odd}}{dt} \propto \frac{1}{Q^8}. \] 
Notice that: 
{\it i)} for the chiral-odd coupling the longitudinal term is suppressed relatively to the transverse one, already at tree level; 
{\it ii)} based on collinear factorization, the chiral-even longitudinal term should be dominating. 
In what follows (Section 4) we show, however, that by taking into account both the GPD crossing properties, along with the corresponding $J^{PC}$ quantum numbers in the $t$-channel, the allowed linear combinations of chiral-even GPDs that  contribute to the longitudinal cross section terms largely cancel each other. As a consequence, the chiral-odd, transverse terms dominate. 

{\em We conclude that spin cannot be disregarded when evaluating the asymptotic trends of the cross section.}  

In addition to assessing the impact of the correct GPD combinations to $\pi^o$ electroproduction, we also developed a model for the hard vertex that takes into account the direct impact of spin through different $J^{PC}$ sequencings.
A model that has been followed is the modified perturbative approach  (\cite{GolKro} and references therein), according to which one has
\begin{eqnarray}
g_{\Lambda_{\gamma^*},\lambda; 0, \lambda^\prime}  =  \int d \tau \int d^2 b \, \hat{{\cal F}}_{\Lambda_{\gamma^*},\lambda; 0, \lambda^\prime}(Q^2,\tau,b) \alpha_S(\mu_R) \exp[-S] \hat{\phi}_\pi(\tau,b) 
\label{collinear}
\end{eqnarray}
where $\hat{{\cal F}}_{\Lambda_{\gamma^*},\lambda; 0, \lambda^\prime}$ is the Fourier transform of the hard (one gluon exchange) kernel, $S$ is the Sudakov form factor,  $\hat{\phi}_\pi$ is the pion distribution amplitude in impact parameter, $b$, space, $\mu_R$ is a renormalization scale. 
In the collinear approximation one obtains the simplified $Q^2$ dependences of Eqs.(\ref{gL_coll},\ref{gT_coll}).    

As we explain in  Section 4, there exist two distinct series  of $J^{PC}$ configurations in the $t$-channel, namely the {\it natural parity} one ($1^{--}, 3^{--} ... $), labeled $V$, and 
the {\it unnatural parity} one ($1^{+-}, 3^{+-} ...$), labeled $A$. We hypothesize that  the two series will generate different contributions to the pion vertex. 
%This statement is equivalent to proposing an alternative to the standard, one gluon exchange mechanism. 
We consider separately the two contributions $\gamma^* (q \bar{q})_V \rightarrow \pi^o$ and 
$\gamma^* (q \bar{q})_A \rightarrow \pi^o$  to the process in Fig.1b.  
What makes the two contributions distinct is that, in the natural parity case (V), the orbital angular momentum, $L$, is the same for the initial and final states, or $\Delta L=0$,
while for unnatural parity (A), $\Delta L =1$. 
We modeled this difference by replacing Eq.(\ref{collinear})  with the following expressions containing a modified kernel 
%%%
\begin{eqnarray}
g^V_{\Lambda_{\gamma^*},\lambda; 0, \lambda^\prime} & =  & \int dx_1 dy_1 \int  d^2 b  
\, \hat{\psi}_V(y_1,b) \, \hat{{\cal F}}_{\Lambda_{\gamma^*},\lambda; 0, \lambda^\prime}(Q^2,x_1,y_1,b) \alpha_S(\mu_R)
 \exp[-S]     \, \hat{\phi}_{\pi^o}(x_1,b)  \nonumber \\
 & = & g_{\pi \, odd}^{V}(Q) \; C^- \\
g^A_{\Lambda_{\gamma^*},\lambda; 0, \lambda^\prime}  & = &  \int dx_1 dy_1 \int d^2  b  
\, \hat{\psi}_A(y_1,b) \, \hat{{\cal F}}_{\Lambda_{\gamma^*},\lambda; 0, \lambda^\prime}(Q^2,x_1,y_1,b) \alpha_S(\mu_R)
\exp[-S]  \, \hat{\phi}_{\pi^o}(x_1,b) \nonumber \\
& = & g_{\pi \, odd}^{A}(Q) \; C^-
\end{eqnarray}
%%%
where, 
\begin{equation}
\hat{\psi}_{A}(y_1,b) = \int d^2 k_T J_1(y_1 b) \psi_V(y_1,k_T) 
\end{equation}
Notice that we now have an additional function, $\hat{\psi}_{V(A)}(y_1,b)$ that takes into account the effect of 
different $L$ states. The higher order Bessel function describes the situation where $L$ is always larger in the initial state. 
In impact parameter space this corresponds to configurations of larger radius. 
The matching of the $V$ and $A$ contributions to the helicity amplitudes is as follows: $f_1, f_4 \propto g^V$, $f_2 \propto g^V+g^A$, 
$f_3 \propto g^V-g^A$, thus explaining the $Q^2$-dependent factors in Eqs.(\ref{f_amps_1}).

In summary, we introduced a mechanism for the $Q^2$ dependence of the process $\gamma^* q \bar{q} \rightarrow \pi^o$, that distinguishes among natural and unnatural parity configurations. We tested the impact of this mechanism using  the modified perturbative approach as a guide. Other schemes could be explored  \cite{Rad}.
  
In the following Section, we validate our approach 
by showing the importance of the $J^{PC}$ structure.  

%%%%% JPC Quantum numbers
\section{Spin, Parity, and Charge Conjugation}
\label{sec4}
Viewed from the $t-$channel the process is $\gamma^* + \pi^0 \rightarrow (u +{\bar u}) \, {\rm or} \, (d + {\bar d}) \rightarrow N + {\bar N}$. 
%whether the quarks are seen as non-local field operators or intermediate states. 
%These are the quantum numbers of the $\rho^0 ,\omega$ mesons and all their orbital excitations, and the $b_1^0, h_1$ mesons and their orbital excitations. 
\subsection{$N {\bar N}$ states}
The $N {\bar N}$ states have well known angular momenta decompositions that we tabulate for reference in Table~\ref{NNbar}.
\begin{table}
\begin{tabular}{c|cccccc}
\hline
  $S / L$          & $0$ & $1$ & $2$ & $3$ & $4$ & $\ldots$ \\
\hline
%\\
$0$ & $0^{-+}$ & $1^{+-}$ & $2^{-+}$ & $3^{+-}$ & $4^{-+}$ & \\
$1$ & $1^{--}$ & $0^{++}$ & $1^{- -}$ & $2^{++}$ & $3^{--}$ &  \\
             &                 & $1^{++}$ &$2^{--}$  & $3^{++}$ & $4^{--}$ & \\
             &                 & $2^{++}$ & $3^{--}$ & $4^{++}$ & $5^{--}$  & \\
\hline
\end{tabular}
\caption{$J^{PC}$ of the $N{\bar N}$ states. }
\label{NNbar}
\end{table}
Before continuing with the t-channel picture it is important to specify clearly what the operators are whose matrix elements are being evaluated between nucleon states,
%As we see above Eq.(\ref{pi_coupling}), for the hard part of a factorized amplitude with the $\pi$ distribution having a Dirac structure $\gamma_5\not\!{q}^\prime$, the hard amplitude has the form of the Dirac matrix $\gamma^+ \gamma^5$. For the $\pi$ distribution having a Dirac structure $ \gamma_5$ alone, the hard amplitude has the form of the Dirac matrix $\sigma^{+ j}\gamma^5$. Then there are two corresponding forms of GPDs involved, those for which the matrix element of the quark correlator going from the s- to t-channel is 
%%%
\begin{eqnarray}
\left \langle p^\prime, \lambda^\prime \mid {\bar \psi}\left(-\frac{z}{2} \right) \Gamma \psi \left(\frac{z}{2} \right) \mid p, \lambda \right\rangle \,  
 \rightarrow \left\langle p^\prime, \lambda^\prime;  \bar{p} \bar{\lambda} \mid {\bar \psi}\left(-\frac{z}{2} \right) \Gamma \psi \left(\frac{z}{2} \right) \mid 0 \right\rangle,
\label{evencor}
\end{eqnarray}
%%%
where for $\pi^o$ electroproduction $\Gamma = \gamma^\mu \gamma_5, \sigma^{+\mu} \gamma_5$. 
To study their correspondence to the t-channel $J^{PC}$ quantum numbers they are expanded into 
an infinite series of local operators which read \cite{JiLebed},
\begin{equation}
O_A^n = {\bar \psi}(0)\gamma^{\{  \mu} i{\overleftrightarrow D}^{ \mu_1} \ldots  i{\overleftrightarrow D}^{ \mu_n \, \}} \gamma^5 \psi(0)
\label{axial_ope}
\end{equation}
\begin{equation}
O_T^n = {\bar \psi}(0)\sigma^{\{ + \mu} i{\overleftrightarrow D}^{ \mu_1}  \ldots i{\overleftrightarrow D}^{ \mu_n \,  \}} \gamma^5 \psi(0)
\label{tensor_ope}
\end{equation}
for the axial-vector, and tensor  cases, respectively (note that tensor corresponds to a C-parity odd operator, while the axial vector has a C-parity even operator). 

The matrix elements of these operator series have different form factor decompositions, and correspondingly different $J^{PC}$ series 
\cite{JiLebed,ChenJi,Hagler,DieIva}. 
In Table \ref{njlsA}  we show the $J^{PC}$ values for the states  $O_A^n \mid 0 \rangle$ along with the corresponding $(S,L)$ values 
for the $N{\bar N}$ states. 
\footnote{There are no $(S,L)$ combinations for $N{\bar N}$ states that can yield the exotic $J^{(-1)^J, \, (-1)^{J+1}}$, so those latter are not connected.}
\begin{table}
\begin{tabular}{c|cccccc}
\hline
%Helicity Flip & No Helicity Flip \\
$n$ & & &  $J^{PC}(S;L)$ \\
\hline
\\
$0$ & $0^{-+}(0;0)$ & $1^{++}(1;1)$  & \\
$1$ & $0^{--}$ &$1^{+ -}(0;1)$ & $2^{- -}(1;2)$ & \\
$2$ & $0^{-+}(0;0)$ &$1^{++}(1;1)$ & $2^{-+}(0;2)$ & $3^{++}(1;3)$ & \\
$3$ & $0^{--}$ & $1^{+ -}(0;1)$ & $2^{- -}(1;2)$ & $3^{+-}(0;3)$ & $4^{--}(1;4)$ & \\
$\ldots$ & &  & $\dots$ & \\
\hline
\end{tabular}
\caption{$J^{PC}$ of the axial operators  with $(S;L)$ for the corresponding $N{\bar N}$ state. Where there are no $(S;L)$ values there are no matching quantum numbers for the $N{\bar N}$ system.}
\label{njlsA}
\end{table}
In Table~\ref{njlsT1}  we show the tensor operators. There are two sets of $J^{PC}$ values, depending on whether $\sigma^{0 j}$ or $\sigma^{3,j}$ are considered,
with opposite $P$ and same $J^C$. For the $\sigma^{+ j}=\sigma^{0 j}+\sigma^{3 j}$ components, the two parity sequences will both be present for each  $J$ and $C$. Considered separately we see that $\sigma^{0 j}$ connects to the $N {\bar N}$ $S=1$ states while $\sigma^{3 j}$ connects to the $N {\bar N}$ $S=0$ and $1$ states. 
\begin{table}
\begin{tabular}{c|cccc|cccc}
\hline
$n $ &   $\sigma^{0j}$ & $J^{PC}(S;L,L^\prime)$ & & & \,\,\, $\sigma^{jk}$ & &  $J^{PC}(S;L)$  \\
\hline
\\
$0$ & $1^{--}(1;0,2)$  & & & & $1^{+-}(0;1)$  \\
$1$ & $1^{- +}$ & $2^{++}(1;1,3)$ & &  &$1^{++}(1;1)$ & $2^{-+}(0;2)$ \\
$2$ & $1^{--}(1;0,2)$ & $2^{+-}$ & $3^{--}(1;2,4)$ & & $1^{+-}(0;1)$ & $2^{--}(1;2)$ & $3^{+-}(0;3)$  \\
$3$ & $1^{-+}$ & $2^{++}(1;1,3)$ & $3^{-+}$ & $4^{++}(1;3,5)$ & $1^{++}(1;1)$ & $2^{-+}(0;2)$ & $3^{++}(1;3)$ & $4^{-+}(0;4)$ \\
$ \ldots $ & & & & $\ldots$ & & & & \ldots
\end{tabular}
\caption{$J^{PC}$ of the tensor operators $\sigma^{0j}$ and $\sigma^{jk}$ with $(S;L)$ for the corresponding $N{\bar N}$ state.}
\label{njlsT1}
\end{table}

%\begin{table}
%\begin{tabular}{c|cccccc}
%\hline
%$n $ & & &$\sigma^{jk}$ $J^{PC}(S;L)$  &\\
%\hline
%\\
%$0$ & $1^{+-}(0;1)$  & \\
%5$1$ & $1^{++}(1;1)$ & $2^{-+}(0;2)$ & \\
%2$ & $1^{+-}(0;1)$ & $2^{--}(1;2)$ & $3^{+-}(0;3)$ & \\
%$3$ & $1^{++}(1;1)$ & $2^{-+}(0;2)$ & $3^{++}(1;3)$ & $4^{-+}(0;4)$ & \\
%$\ldots$ & & & \dots & \\
%\end{tabular}
%\caption{$J^{PC}$ of the tensor operators $\sigma^{jk}$ with $(S;L)$ for the corresponding $N{\bar N}$ state.}
%\label{njlsT2}
%\end{table}
\subsection{$\gamma^* \pi^o$ states}
The $\gamma^* + \pi^0$ state must be C-parity negative. The lowest $J^{PC}$ series that can be involved will be ($1^{- -}, 2^{- -}, 3^{- -},  \ldots$) and ($1^{+ -} , 2^{+ -}, 3^{+ -}, \ldots $). 
%Now for $\pi^o \gamma^*$ 
For helicity 0, or longitudinal $\gamma^*$, the coupling to the t-channel is in an eigenstate of C-parity.  That requires $L$-excitations to be even, since $C=(-1)^{L+1}$. 
For the transverse $\gamma^*$ the  allowed $J^{PC}$ values can have either $L$-even or odd, since C is not an eigenvalue then. 
Forming symmetric or antisymmetric combinations of the two transverse states selects the $C$-parity. So allowed quantum numbers are $1^{--}$ with all L excitations and $1^{+-}$ with even L excitations to guarantee $ C=- $. That is $J^{\pm \, -}=(L-1, L, L+1)^{\pm \, - }$. These are collected in Table~\ref{gammapi}.
\begin{table}
\begin{tabular}{c|c|c|c|c|cc}
\hline
            & $L=0$ & $1$ & $2$ & $3$  & $4$ & $\ldots$ \\
\hline
%\\
$\Lambda_\gamma=0$         & $1^{+-}$ & & $1^{+-},2^{+-},3^{+-}$ &   & $3^{+-},4^{+-},5^{+-}$ \\
$\mid \Lambda_\gamma\mid= 1$  & $1^{+ -}$ & $0^{- -},1^{- -},2^{- -}$ & $1^{+-},2^{+-},3^{+ -}$ & $2^{- -},3^{- -},4^{- -}$ & $3^{+-},4^{+-},5^{+-}$ &  \\
%             &                 & $1^{++}$ &$2^{--}$  & $3^{++}$ & $4^{--}$ & \\
 %            &                 & $2^{++}$ & $3^{--}$ & $4^{++}$ & $5^{--}$  & \\
\hline
\end{tabular}
\caption{$J^{PC}$ of the $\gamma^* \pi^0$ states. }
\label{gammapi}
\end{table}
%%%%%%%%%% Not completely happy with this Table yet . . . .

%Note that for the $\mid t/s \mid <<1$ $J$ odd is leading order compared to the even values. This applies for all couplings to the GPDs.
We see in Table~\ref{gammapi} that ($1^{+ -}, 3^{+ -}, \ldots$) will be candidates for one series of couplings, corresponding to the  $S=0, J =L$ sequence in Table \ref{NNbar}. Analogously, the ($2^{- -}, 4^{- -}, \ldots$) series corresponds to $S=1, J=L$ for the $N {\bar N}$ couplings. The ($1^{+ -}, 3^{+ -}, \ldots$) $J^{PC}$ values occur in the tensor, chiral odd case (Table~\ref{njlsT1}),
while the ($2^{- -}, 4^{- -}, \ldots$) series occurs in both the axial, chiral even case (Table~\ref{njlsA}) and the tensor case (Table~\ref{njlsT1}). 

%%%%%% Continue here with the relations to the GPDs like H(x...) - H(-x...) This is done by comparing the FFs associated with each n to those in the polynomial decomposition into powers of \xi. See HAegler paper and also Diehl's also
\subsection{Connection to GPDs}
How are these n-operators connected to the GPDs? In Ref.~\cite{JiLebed} the basic scheme is explained. It is a natural generalization to off-forward scattering of the Mellin transform method for PDFs.  From the connection between Mellin moments and $J^{PC}$ quantum numbers it can be shown how the GPDs can be decomposed into sequences of $t-$channel quantum numbers. For the chiral even vector and axial vector case these relations~\cite{DieIva} are reproduced in Table~\ref{VAGPD}. The corresponding tensor, chiral odd case has not been decomposed previously. We derive this case in the following and display the results in Table~\ref{TTGPD}.

\begin{table}
\begin{tabular}{c|cccccc}
\hline
${\rm Chiral \, \, Even} \, \,  GPD $ &  $J^{PC}$  &\\
\hline
\\
$H(x,\xi,t)-H(-x,\xi,t)$ & $0^{++}$,  $2^{++}$,  $\ldots$ &   & $(S=1)$ &    &\\
$E(x,\xi,t)-E(-x,\xi,t)$ & $0^{++}$, $2^{++}$, $\ldots$ &  & $(S=1)$ \\
$\widetilde{H}(x,\xi,t)+\widetilde{H}(-x,\xi,t)$ & $1^{++}$, $3^{++}$, $\ldots$ & &  $(S=1)$ \\
$\widetilde{E}(x,\xi,t)+\widetilde{E}(-x,\xi,t)$ & $0^{-+}$, $1^{++}$, $2^{-+}$, $3^{++}$, $\ldots$ & & $(S=0,1)$ \\
\hline
\\
$H(x,\xi,t)+H(-x,\xi,t)$ & $1^{--}$, $3^{--}$, $\ldots$ & &   $(S=1)$ \\
$E(x,\xi,t)+E(-x,\xi,t)$ & $1^{--}$, $3^{--}$, $\ldots$ & &   $(S=1)$ \\
$\widetilde{H}(x,\xi,t)-\widetilde{H}(-x,\xi,t)$ & $2^{--}$, $4^{--}$, $\ldots$ & &  $(S=1)$ \\
$\widetilde{E}(x,\xi,t)-\widetilde{E}(-x,\xi,t)$ & $1^{+-}$, $2^{--}$, $3^{+-}$, $4^{--}$,  $\ldots$ & & $(S=0,1)$ \\
\hline
%$ \ldots $ & & & $\ldots$ &
\end{tabular}
\caption{$J^{PC}$ decompositions for the chiral even GPDs. The C-parity even (odd) combinations are in the upper (lower) section.}
\label{VAGPD}
\end{table}

\subsubsection{Axial GPDs}
The combinations of axial GPDs in Table~\ref{VAGPD} can be compared with the $J^{PC}$ values accessed by $\pi^0$ production in Table~\ref{gammapi}. 

By analogy with the formalism of parton distributions, 
we define $F_q(x,\xi,t)$ ($F=H,E$) in the interval $-1 \leq x \leq 1$, with the following identification of anti-quarks,
\begin{eqnarray}
%F_q(x,\xi) & =  &  F_q(x,\xi)  \; \; \; x \geq  0 \\
F_{\bar{q}}(x,\xi,t) & = & -  F_q(x,\xi,t)    \; \; \;  x<0.                            
\end{eqnarray}
From this expression one defines 
\begin{eqnarray}
F_q^- & = & F_q(x,\xi,t) +  F_q(-x,\xi,t) \\
F_q^+ & = & F_q(x,\xi,t) -F_q(-x,\xi,t) , 
\end{eqnarray} 
where $F_q^-$ is identified with the  flavor non singlet,  valence quarks distributions, and $\sum_{q} F_q^+$ with the flavor singlet, sea quarks distributions. 
Similarly,
\begin{eqnarray}
\widetilde{F}_q^- & = & \widetilde{F}_q(x,\xi,t) -  \widetilde{F}_q(-x,\xi,t) \\
\widetilde{F}_q^+ & = & \widetilde{F}_q(x,\xi,t) + \widetilde{F}_q(-x,\xi,t). 
\end{eqnarray}
The C-parity odd values that connect with $\gamma^* \pi^o$ are in $\widetilde{H}^-$ and $\widetilde{E}^-$. In the former, contributions begin with $2^{--}$. This is the lowest of the non-singlet, chiral even $t$-channel contributions to the $\pi^0$ production. Now $\widetilde{H}(x,0,0) = g_1(x)$ contributes significantly to the longitudinal asymmetry in the nucleon PDFs. For the process here, however, the $J^{PC}=2^{--}$ will suppress the non-singlet contribution for small $x$ and $\mid t \mid$. The considerably smaller value of the longitudinal cross section for $\pi^0$ corroborates this conclusion. The $\widetilde{E}$ is known to receive a large contribution from the $\pi$ pole, although that pole does not enter the neutral $\pi$ production. After the pion pole is removed the values of $\widetilde{E}$ are expected to be much smaller. These observations suggest the importance of the chiral odd contributions.

\subsubsection{Tensor GPDs}
We continue with the same logic for the chiral odd decompositions in Table~\ref{TTGPD}. 
The two sequences of $J^{- -}$ and $J^{+-}$ contribute equally to the chiral odd GPDs and couple to the $\pi^0$ production. These sequences correspond to the $\rho, \, \omega$ (vector) and $b_1, \, h_1$ (axial-vector) mesons with even $L$ excitations. The decomposition into the $t$-channel of the matrix elements of the non-local quark correlators into the two distinct series of local operators lead us to distinguish two different dependences on $Q^2$ of the $\pi^o$ coupling, $g_{\pi \, odd}^V(Q)$ ($\rho, \, \omega$ series, lower left in Table \ref{TTGPD}, and  $g_{\pi \, odd}^A(Q)$ ($b_1, \, h_1$ series, lower right in Table \ref{TTGPD}) as discussed in Section \ref{sec3}. 

\section{Conclusions}
\label{sec5}
Exclusive pseudo-scalar meson electroproduction is directly sensitive to leading twist chiral-odd GPDs which can at present be extracted from available data in the multi-GeV region. We examined and justified this proposition further by carrying out a careful analysis of some of the possibly controversial issues that had arisen.  
After going through a step by step derivation of  the connection of the helicity amplitudes formalism with the cartesian basis, we showed how the dominance of the chiral-odd process follows unequivocally owing to the values allowed for the $t$-channel spin, parity, charge conjugation and from the GPDs crossing symmetry properties.
This observation has important consequences for the the $Q^2$-dependence of the process. Our $J^{PC}$ analysis supports 
the separation between the $Q^2$ dependence of the photo-induced transition functions 
for the even and odd parity combinations into $\pi^0$ thus
reinforcing the idea that spin related observables exhibit a non trivial asymptotic behavior. Finally, in an effort to streamline the otherwise cumbersome multi-variable dependent structure functions, we presented simplified formulae displaying leading order contributions of the GPDs and their $Q^2$ dependent multiplicative factors. 
A separation of the various chiral-odd GPDs contributions can be carried out provided an approach that allows to appropriately fix their parameters and normalizations is adopted as the one we presented here.

\begin{table}
\begin{tabular}{c|lcc|cccc}
\hline
${\rm Chiral \, \, Odd} \, \,  GPD $ &  $J^{- \, C}$ && & $J^{+ \, C}$\\
\hline
\\
$H_T(x,\xi,t)-H_T(-x,\xi,t)$ & $2^{-+}$,  $4^{-+}$,  $\ldots$ & $(S=0)$  & &$1^{++}$, $3^{++}$ \ldots & $(S=1)$ & &\\
$E_T(x,\xi,t)-E_T(-x,\xi,t)$ & $2^{-+}$,  $4^{-+}$,  $\ldots$ & $(S=0)$  & &$1^{++}$, $3^{++}$ \ldots & $(S=1)$ & &\\$\widetilde{H}_T(x,\xi,t)-\widetilde{H}_T(-x,\xi,t)$ & & & & $1^{++}$, $3^{++}$, $\ldots$ &  $(S=1)$ \\
$\widetilde{E}_T(x,\xi,t)-\widetilde{E}_T(-x,\xi,t)$ & $2^{-+}$,  $4^{-+}$,  $\ldots$ & $(S=0)$  & &$3^{++}$, $5^{++}$ \ldots & $(S=1)$ & &\\
\hline
\\
$H_T(x,\xi,t)+H_T(-x,\xi,t)$ & $1^{--}$, $2^{--}$, $3^{- -}$ $\ldots$ & &   $(S=1)$ & $1^{+-}$, $3^{+-}$ \ldots & (S=0)\\
$E_T(x,\xi,t)+E_T(-x,\xi,t)$ & $1^{--}$, $2^{--}$, $3^{- -}$ $\ldots$ & &   $(S=1)$ & $1^{+-}$, $3^{+-}$ \ldots & (S=0)\\
$\widetilde{H}_T(x,\xi,t)+\widetilde{H}_T(-x,\xi,t)$ & $1^{--}$, $2^{--}$, $3^{- -}$ $\ldots$ & &   $(S=1)$ & \\
$\widetilde{E}_T(x,\xi,t)+\widetilde{E}_T(-x,\xi,t)$ &$2^{--}$, $3^{--}$, $4^{- -}$ $\ldots$ & &   $(S=1)$ & $3^{+-}$, $5^{+-}$ \ldots & (S=0)\\
\hline
%$ \ldots $ & & & $\ldots$ &
\end{tabular}
\caption{$J^{PC}$ decompositions for the chiral odd GPDs. The C-parity even (odd) combinations are in the upper (lower) section.}
\label{TTGPD}
\end{table}

\vspace{0.5cm}
We thank P. Kroll, V. Kubarovsky, M. Murray, A. Radyushkin, P. Stoler, C. Weiss, for many interesting discussions and constructive remarks.  Graphs were made using Jaxodraw \cite{jaxo}. This work was supported by the U.S. Department
of Energy grants DE-FG02-01ER4120 (J.O.G.H., S.L.),  DE-FG02-92ER40702  (G.R.G.).

%%%%%%%%%%%%%%%%%%%%%%%%%%%%%%%%%%%%%%%%%%%%%%%%%%%%%%%%%%%%%%%%

\end{document}